\documentclass[aps,prc,twocolumn,floatfix,showpacs,a4paper,
nofootinbib,amsmath,amssymb]{revtex4-1}
\usepackage{graphicx}
\usepackage{dcolumn}
\usepackage{bm}
\usepackage{color}

\newcommand{\be}{\begin{equation}}
\newcommand{\ee}{\end{equation}}
\newcommand{\ba}{\begin{eqnarray}}
\newcommand{\ea}{\end{eqnarray}}
\newcommand{\bd}{\begin{displaymath}}
\newcommand{\ed}{\end{displaymath}}
\renewcommand{\vec}[1]{\mbox{\boldmath$#1$}}

\begin{document}
\title{Differential Hanbury-Brown-Twiss for an exact hydrodynamic model with rotation}

\author{S. Velle and L.P. Csernai}
\affiliation{Institute of Physics and Technology, University of Bergen,
Allegaten 55, 5007 Bergen, Norway }
\date{\today}
\begin{abstract}
We study an exact rotating and expanding solution of the fluid dynamical
model of heavy ion reactions, that take into account the rate of slowing
down of the rotation due to the longitudinal and transverse
expansion of the system. The parameters of the
model are set on the basis of realistic 3+1D fluid dynamical
calculation at TeV energies, where the rotation
is enhanced by the build up of the Kelvin Helmholtz Instability
in the flow.
\end{abstract}


\maketitle

\section{Introduction}

In peripheral heavy ion collisions the participant system has
large angular momentum.
In hydrodynamical model calculations it has been shown that
this shear leads to a significant vorticity \cite{CMW13,Vov14}.
When Quark-Gluon Plasma
(QGP) is formed with low viscosity \cite{CKM,KSS,NNG},
new phenomena may occur like rotation
\cite{hydro1}, or turbulence,
which shows up in form of a starting Kelvin-Helmholtz
Instability (KHI) \cite{hydro2,YH,WNC13}.
It was also shown \cite{CF} that in peripheral collisions even if the
shear flow is neglected and the same boost invariant longitudinal
velocity is assumed at all transverse points, in the Color Glas Condensate
(CGC) model initial transverse flow develops and it contributes to angular
momentum in the same direction as the angular momentum arising from 
the target and projectile motion in the participant system. 

Rotation in heavy ion collision has only recently been considered, 
our aim is to detect it using two particle correlation. 

The Differential Hanbury Brown and Twiss (DHBT) method 
has been introduced earlier in \cite{DCF}. 
Previously the method has been applied to a high resolution 
Particle in Cell Relativistic (PICR) fluid dynamical model \cite{DCF2}. 
Here we will look at how the DHBT can be used for the exact 
hydro model \cite{CsoN14,ReHM}.
Rotation in exact models have been investigated as well in \cite{HNX}.

We look at the values from the exact hydro model and 
determine the effect rotation has
on the correlation functions (CF) for detectors at different positions.

\section{Correlation function for exact hydro model.}

The two particle correlation function for this model is found with the same 
method used in \cite{DCF} where the source function is

\be
S(x,k)=\frac{n(x)k^\mu \sigma_\mu}{C_n}\exp
\left[-\frac{k^\mu u_\mu}{T(x)}\right] \ ,
\label{sf}
\ee 

where $C_n$ is a constant, $k^\mu$ is the average 4 vector 
momentum of two pions, $k=(p_1+p_2)/2$ and the momentum difference 
is $q=(p_2-p_1)$, $u_\mu$ is the 4-vector velocity of the source, 
$\sigma_\mu$ is the normal of the freeze out hypersurface and 
$T(x)$ is the temperature distribution.
The density for the exact hydro model \cite{ReHM} is given by 
\be
N(r_\rho,r_y)=N_B \frac{C_n}{V}\exp(-r^2_\rho/2R^2)\exp(-r^2_y/2Y^2)
\ee
or using the scaling variables in the 
out ($\rho,\ R$), 
side ($\varphi,\ \Theta$), 
long ($y,\ Y$) directions
\be
N(s_\rho,s_y)=N_B \frac{C_n}{V}\exp(-s_\rho/2)\exp(-s_y/2) ,
\label{density}
\ee
where $s_\rho = r^2_\rho / R^2$ and
      $s_y    = r^2_y / Y^2$.

We use the finite size cylindrical shape source as described in eq. (10) of \cite{ReHM}

\be
\int_0^{\infty}\!\!\! \int_{-\infty}^{\infty} \int_0^{2\pi} \!\!\! 
r_\rho dr_\rho dr_y d\varphi = 
R^2 Y\!\! \int_0^1 \!\! \int_0^1 \!\! \int_0^{2\pi} 
\frac{ds_y ds_\rho d\varphi}{\sqrt{s_y}} ,
\label{vol}
\ee 
and the  integral $J(k,q)$ for this model will be
\be
\begin{split}
J(k,q) \propto & \int_0^1 \int_0^1 \int_0^{2\pi} w_s 
\gamma_s \left(k_0 + \vec{k} \cdot \vec{v_s}\right) \times \\ 
& \exp\left[-\frac{\gamma_s}{T_s}((k_0+q_0/2)-(\vec{k} +
\vec{q}/2)\cdot \vec{v_s}) \right] \times \\ 
& \exp(i\vec{q} \cdot \vec{x}) e^{-s_\rho/2}e^{-s_y/2} 
\frac{ds_y ds_\rho d\varphi}{\sqrt{(s_y)}}   \ ,
\end{split}
\label{Jkq}
\ee
where $k_0 = \sqrt{\frac{2m_\pi}{\hbar c}+k^2}$ and 
$q_0=\dfrac{\vec{k} \cdot \vec{q}}{k_0}$, 
$w_s$ is a weight function, $w_s \propto k^\mu \sigma_\mu$, 
and the temperature profile is flat with a value of 250 MeV.

We have the single particle distribution integral
\be
\begin{split}
& \int d^4x S(x,k) \propto \int w_s \gamma_s 
\left(k_0 + \vec{k} \cdot \vec{v_s}\right) \times \\ 
& \exp\left[-\frac{\gamma_s}{T_s}(k_0-\vec{k} \cdot \vec{v_s}) \right] 
 e^{-s_\rho/2}e^{-s_y/2} 
\frac{ds_y ds_\rho d\varphi}{\sqrt{(s_y)}} \ ,
\end{split}
\label{spd}
\ee
and the correlation function is given by \cite{DCF} 
\be
C(k,q) = 1+ \frac{Re[J(k,q)J(k,-q)]}{\left|\int d^4x S(x,k) \right|^2}
\label{cf}
\ee
so the constants outside the integrals will cancel.

\bigskip

The velocity consists of a radial expansion in the out 
direction, an angular velocity, 
where the rotation is in the reaction plane and also 
an axis-directed expansion in the side direction.
The velocity in the out ($\rho$), side ($\varphi$),
long-direction ($y$) is then
\be
 \vec v_s =\left(\dot{R}\sqrt{s_\rho},
   R \omega \sqrt{s_\rho} ,\dot{Y} \sqrt{s_y} \right)\ ,
\label{vs-osl}
\ee
or in the $x,\ y,\ z$ directions where $x$ is the 
direction of the impact parameter,
$y$ is the longitudinal (rotation) axis and 
$z$ is the (beam) collision axis.
\be
\begin{split}
& \vec v_s = \left(
\dot{R} \sqrt{s_\rho}\ \sin(\varphi) + 
R\omega \sqrt{s_\rho}\ \cos(\varphi), \right. \\ 
& \left. \dot{Y} \sqrt{s_y},
\dot{R} \sqrt{s_\rho}\ \cos(\varphi) - 
R\omega \sqrt{s_\rho}\ \sin(\varphi) \right)\ .
\end{split}
\label{vel}
\ee

\begin{widetext}

\begin{figure}[ht]  
\begin{center}
\resizebox{0.34\columnwidth}{!}
{\includegraphics{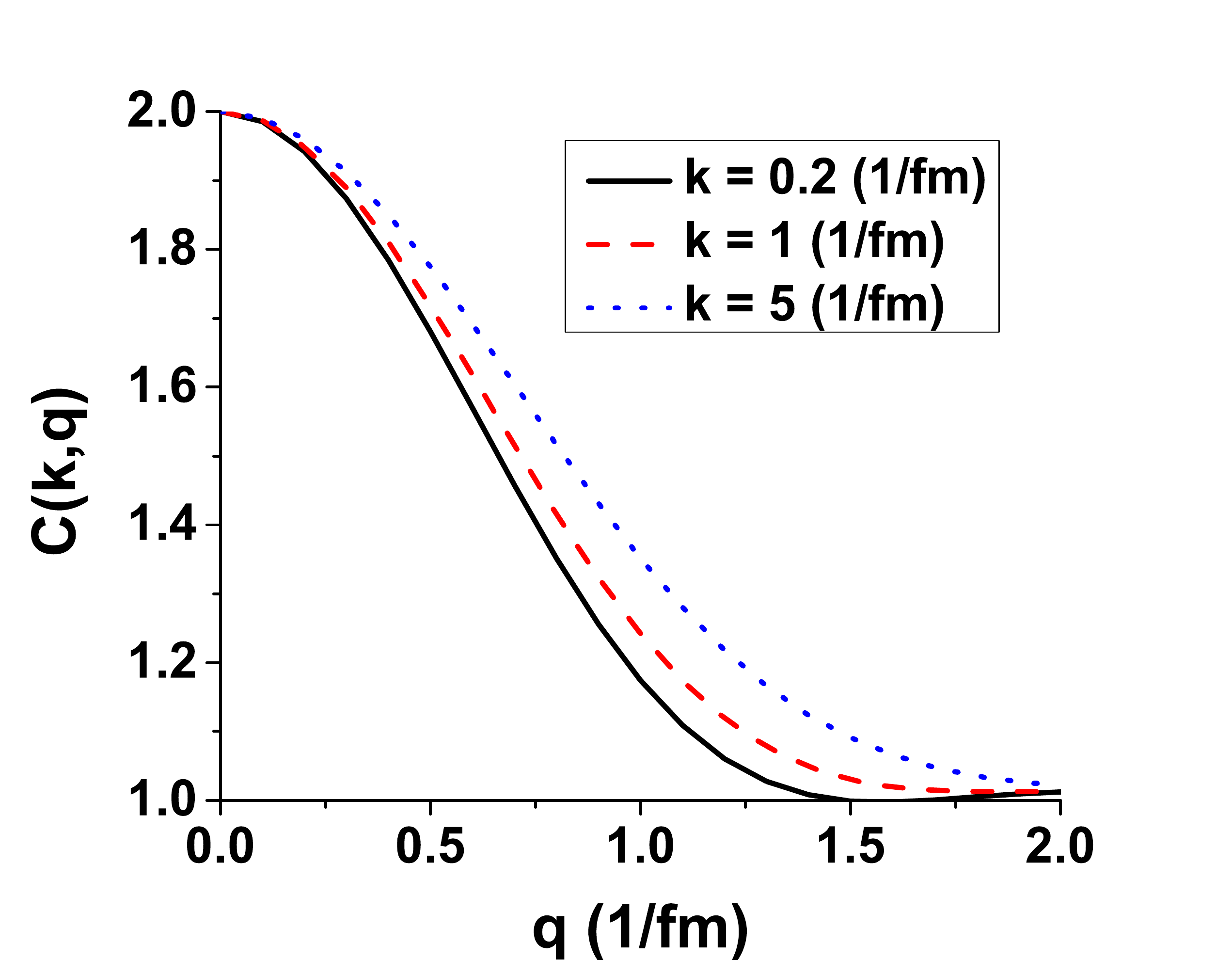}}\!\!\!\!\!\!
\resizebox{0.34\columnwidth}{!}
{\includegraphics{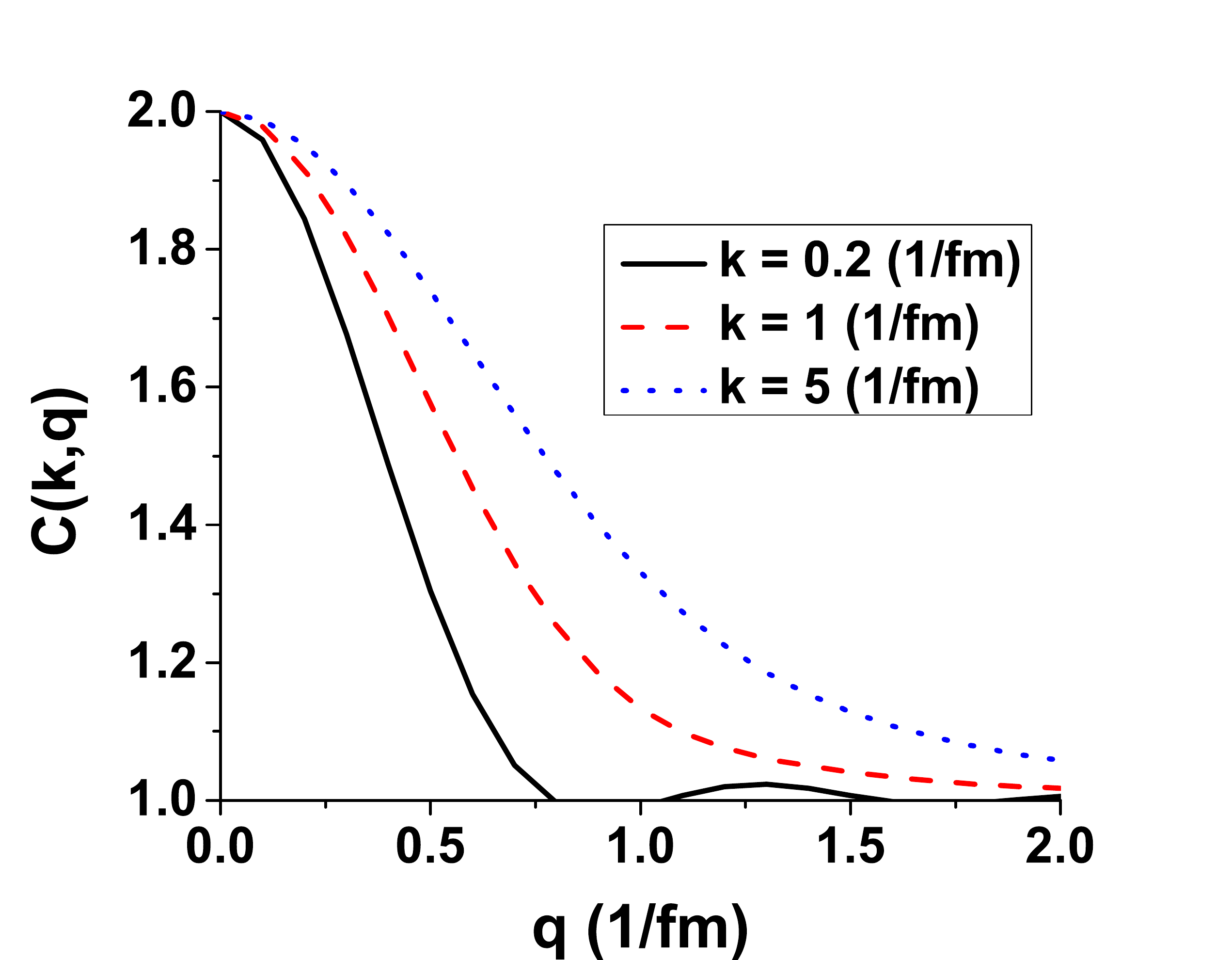}}\!\!\!\!\!\!
\resizebox{0.34\columnwidth}{!}
{\includegraphics{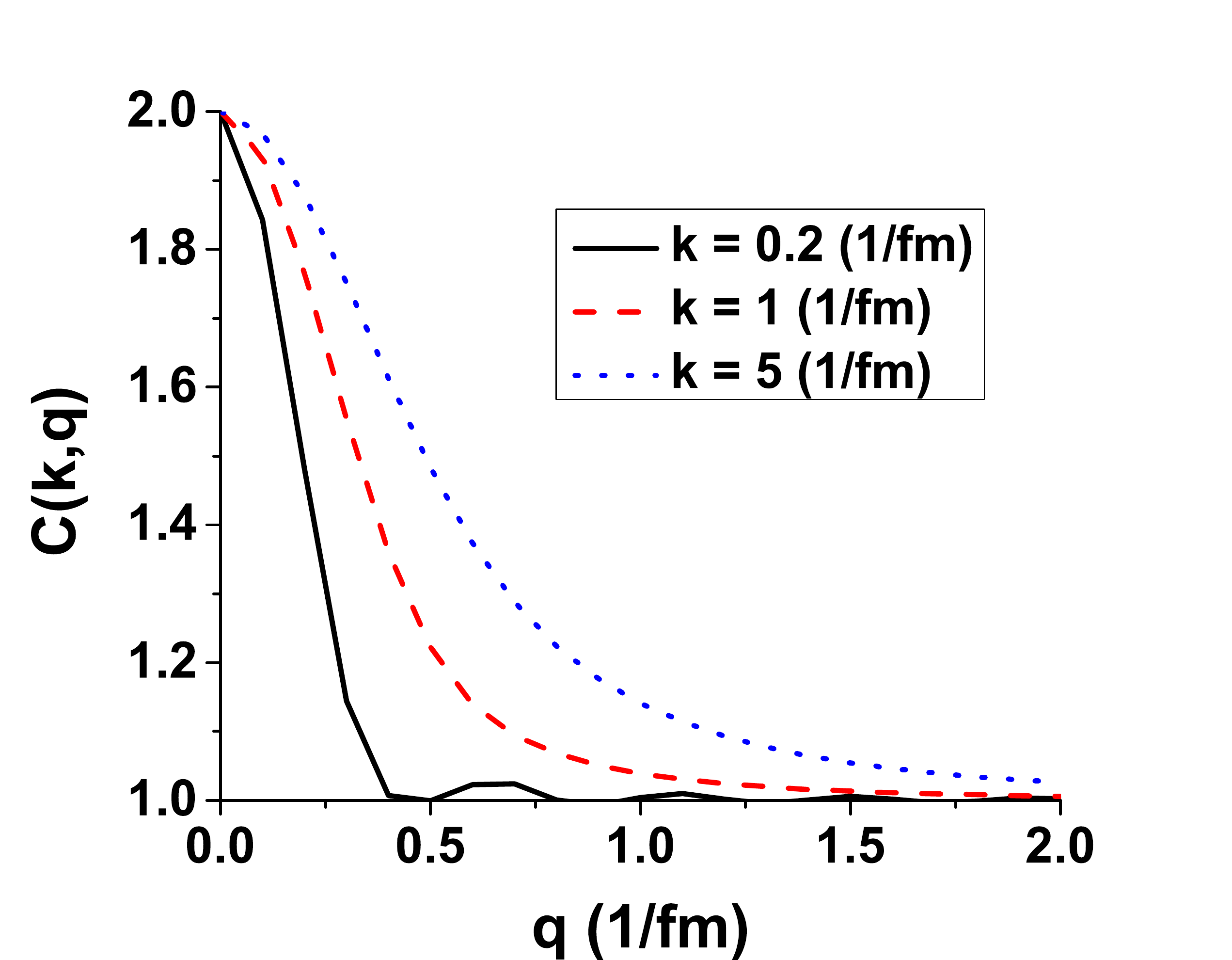}}
\end{center}
\caption{ (color online)
Correlation Function $C(k,q_{out})$ for the exact hydro model 
as function of $q=q_{out}$, with  \\
(a) R = 2.500 fm, $\dot{R}$ = 0.250 c, Y = 4.000 fm, 
$\dot{Y}$ = 0.300 fm, $\omega$ = 0.150 c/fm,
at t = 0.0 fm/c. (left figure) \\
(b) R = 3.970 fm, $\dot{R}$ = 0.646 c, Y = 5.258 fm, $\dot{Y}$ = 0.503 fm, 
$\omega$ = 0.059 c/fm, 
at t = 3.0 fm/c. (middle figure) \\
(c) R = 7.629 fm, $\dot{R}$ = 0.779 c, Y = 8.049 fm, $\dot{Y}$ = 0.591 fm, 
$\omega$ = 0.016 c/fm, 
at t = 8.0 fm/c. (right figure) \\
The solid black line is for $k=0.2 \, fm^{-1}$, the 
dashed red line is for $k=1 \, fm^{-1}$ 
and the dotted blue line is for $k=5 \, fm^{-1}$.
}
\label{ckqt}
\end{figure} 

\end{widetext}

The average transverse radius is $R=\sqrt{XZ}$, and we use this 
value when the exact model is studied.
Using the values from Table~\ref{t1} we show the correlation 
function as function of $q=q_{out}$ in  Figures 
\ref{ckqt} $a$ and $b$ for two different configurations.

As the system size increases with time we get a narrower
distribution in $q_{out}$ for the correlation function. 
We also notice a wider distribution for increasing values of $k$.

\section{Results - Differential HBT for exact hydro model.}

Let us now "Event by Event" evaluate two correlation functions at two
different k-vectors in the plane of rotation.
The differential correlation function \cite{DCF} is obtained 
by taking the difference between
the correlation functions with e.g. at detector positions 
$\vec k^+ = (a,0,b) k$ 
and subtracting the CF for a detector at 
$\vec k^- = (a,0,-b) k$  $(a^2+b^2=1)$ in $x,y,z$ coordinates,
\be
\Delta C(k,q) \equiv C(\vec k^+ ,\vec q) - C(\vec k^-,\vec q) \ .
\label{dcf}
\ee
The integrals for 
this model cannot be given in analytic form, so the CF 
and DCF need to be integrated numerically.
The momentum difference vector $\vec q$ may point in different
directions, and it is usual to use the 
$q_{out},\ q_{long},\ q_{side}$ system of directions. 
In the present work we show the $q_{out}$ dependence
of the Correlation functions only, where $\vec k \parallel \vec q_{out}$.
\bigskip

{\bf Detectors}

\begin{figure}[ht] 
\begin{center}
\includegraphics[width=0.95\columnwidth]{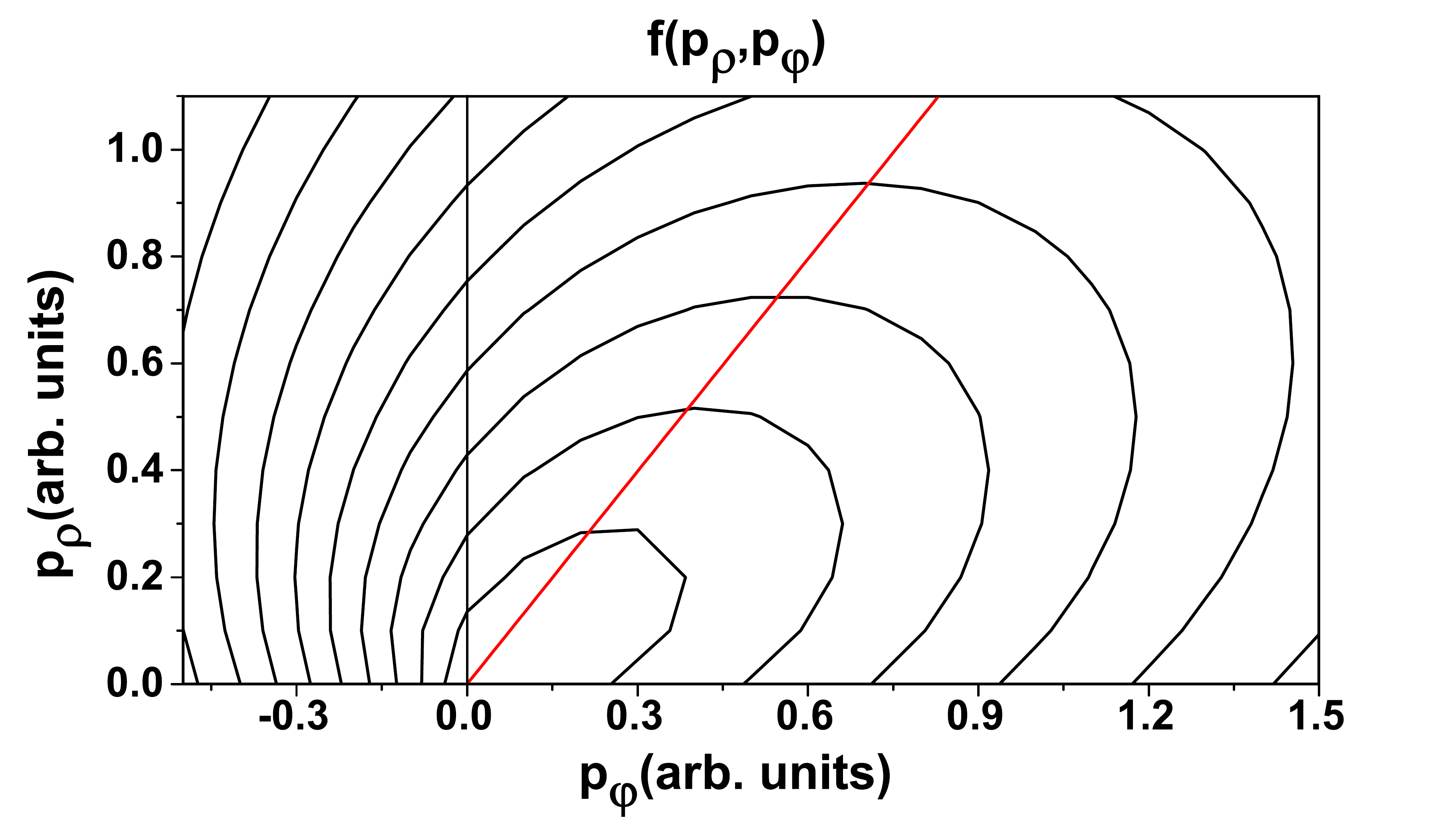}
\end{center}
\vskip -4mm
\caption{ (color online)
The schematic phase space distribution
of the rotating and expanding source in the momentum
space. The momentum of the expansion increases with the radius
just as the momentum arising from the rotation. 
So, higher radial flow momenta correspond to higher rotation momenta also, as given
by eq.(\ref{vs-osl}) and indicated by the red line. 
At constant $p_\rho$ the distribution peaks at the momentum $p_\phi$
indicated by the red line; the J\"uttner distribution, eq.(\ref{vs-osl}), 
is not symmetric, it is elongated towards higher momenta, see Sec. 2.4.2 of 
\cite{LPS}.
The resulting thermally smeared distribution is indicated by the
contour lines.}
\label{sketch}
\end{figure}

As we can see from eq. (8) the rotation leads to an asymmetry, both
the $\rho$ and $\varphi$ components of the velocity $\vec v_s$ depend
linearly on  $\sqrt{s_\rho}$, thus the flow velocities of the
system at a given FO time have a velocity profile
$v_s^{(\varphi)} = {\rm const. } v_s^{(\rho)}$.
The corresponding momentum of the fluid motion has
approximately the same characteristics, and this is indicated
by the red full line in Fig. \ref{sketch}. This characteristic flow velocity
distribution determines the final momentum distribution of the
emitted particles where the contribution of a fluid element at
a radial coordinate $s_\rho$ leads to a tangential momentum
distribution centered around
$v_s^{(\varphi)}= R \omega \sqrt{s_\rho}$.
The resulting momentum distribution of the emitted particles
from different radii are sketched in Fig.
\ref{sketch}.

\begin{table}[ht]
\begin{tabular}{ccccccc} \hline\hline \phantom{\Large $^|_|$}
	$t$  & $Y$ & $\dot{Y}$ & $\omega$ & $R$ & $\dot{R}$ & $\varphi$ \\
	(fm/c) & (fm) & (c)   &  (c/fm) &  (fm)   &   (c)    & (Rad) \\
	\hline
	0.   & 4.000   & 0.300 & 0.150   &  2.500   &  0.250    & 0.000 \\
	3.   & 5.258   & 0.503 & 0.059   &  3.970   &  0.646    & 0.307 \\
	8.   & 8.049   & 0.591 & 0.016   &  7.629   &  0.779    & 0.467 \\
\hline
\end{tabular}
\caption{
	Time dependence of some characteristic parameters of the
	fluid dynamical calculation presented in ref. \cite{ReHM}.  R is
	the average transverse radius, Y is the longitudinal length of the
	participant system, $\varphi$ is the angle of the rotation
        of the interior region of the system,
        around the $y$-axis, measured from the horizontal,
	beam ($z$) direction in the reaction, $[x,z]$ plane,
	$\dot{R},\ \dot{Y}$ are the speeds of expansion in transverse and
	longitudinal directions, and $\omega$ is the angular velocity of the
	internal region of the matter during the collision.
}
\label{t1}
\end{table}

Although our spatial source configuration is azimuthally symmetric,
our phase-space configuration is not, because of the given direction of the
rotation. As eq. (\ref{Jkq}) shows the CF depends on  $\vec q \cdot \vec v_s$
and  $\vec q \cdot \vec x$, thus reversing the direction of either 
$\vec q$ or $\vec v_s$ will change the CF! For vanishing rotation
the CF would be the same for different azimuthal angles (e.g. $d_z = \pm b$)
if the spatial source is azimuthally symmetric.  If there is no 
expansion the  rotation does not change the CF integrals either.

We use a detector placed at $(x,y,z)=(0.935,0,0.353)$, so 
that $\vec{k} = (0.935,0,0.353)k$ and 
$\vec q = \vec q_{out}=(0.935,0,0.353)q_{out}$,
so that both $\vec k$ and $\vec q$ are parallel, are in the reaction 
plane and are orthogonal to the rotation axis, $\vec y$.

If we look at the azimuthal integrals for $\vec k \cdot\vec v_s$
for the detectors at $\vec{k} = 
\left(ak,0,\pm bk\right)$ in eqs. (\ref{Jkq},\ref{spd}),
we have the integrals below, which will have a non-zero difference 
for $a \neq b$, $\dot{R} \neq 0$ and $\omega \neq 0$, both for $\vec k$
and $\vec q_{out}$:
\ba
\int_0^{2\pi} \!\!\!
\exp\left(k \sqrt{s_\rho}\
 \left(  a [\dot{R} \sin(\varphi) + R \omega \cos(\varphi)]\right.\right. 
\nonumber \\
\left.\left. + c [\dot{R} \cos(\varphi) - R \omega \sin(\varphi)]\right)\right) d\varphi 
\nonumber\\
\neq \int_0^{2\pi} \!\!\!
\exp\left(k \sqrt{s_\rho}\
 \left(    a [\dot{R} \sin(\varphi) + R \omega \cos(\varphi)]  \right. \right.
\nonumber\\
\left.\left. - c [\dot{R} \cos(\varphi) - R \omega \sin(\varphi)]\right)\right) d\varphi
\ \ .
\label{integral2}
\ea
This is so because, although the spatial distribution does not depend on 
$\varphi$, the momentum space distribution depends on $v_\varphi$.

In the case of a realistic dynamical configuration, especially in 
the initial configuration, the local rotation velocity component from the shear flow is 
maximal at higher distances in the $\pm x$ directions and pointing towards the
$\pm z$ direction.

\bigskip
\bigskip
\bigskip
For different detector positions we would expect different 
values of the correlation function, and ideally we would 
place them along  or near the direction of highest speed 
in the side direction which would usually be in the 
beam direction.

In the exact model discussed here
we can see that at small values of $k$ there 
is little to no difference in the DCF,
but the difference will increase for higher values. 
As the system grows in size the distribution becomes more 
narrow for the CF and the DCF will become smaller for larger 
values of the relative momentum q.

\begin{widetext}

\begin{figure}[ht]  
\begin{center}
\resizebox{0.34\columnwidth}{!}
{\includegraphics{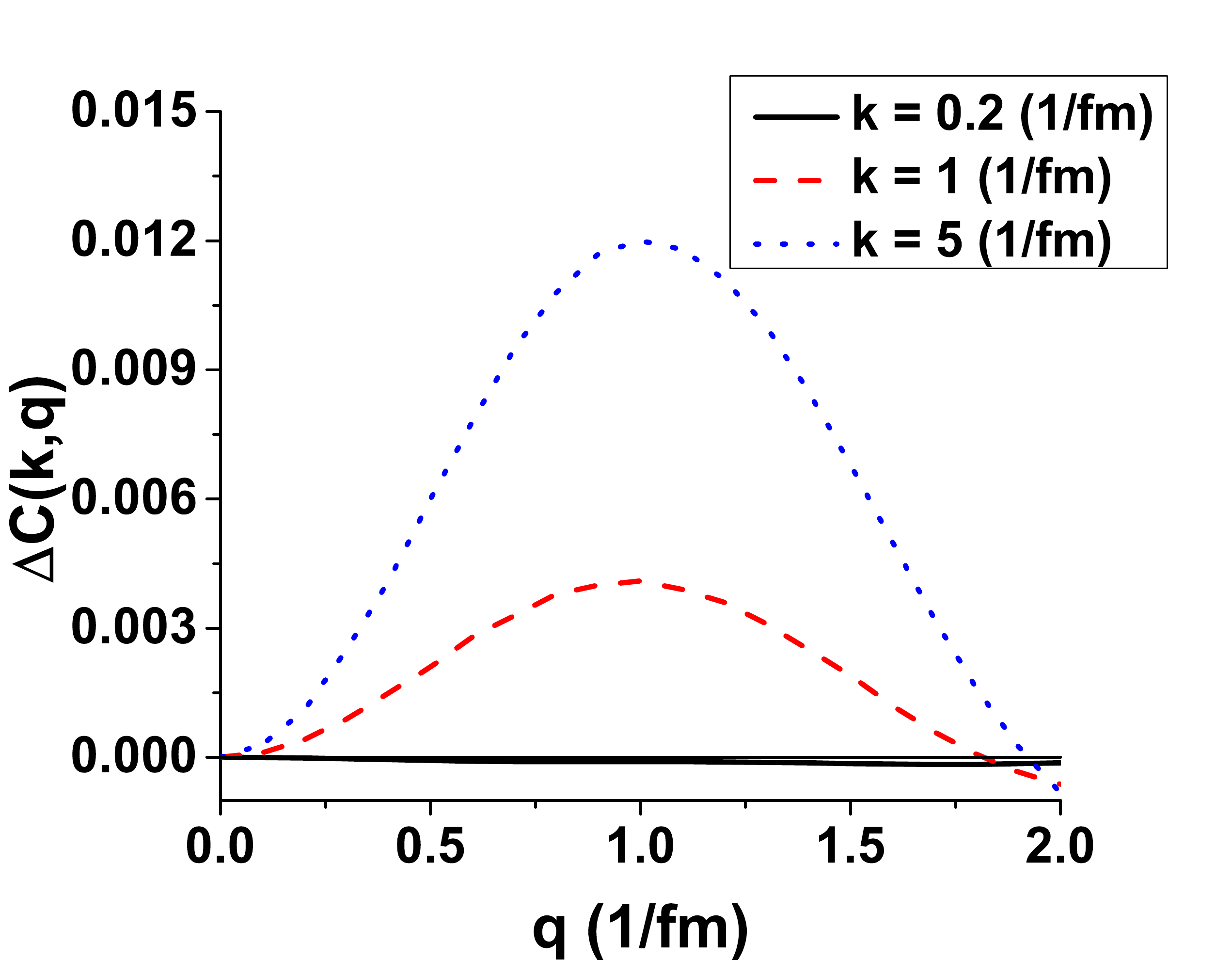}}\!\!\!\!\!\!
\resizebox{0.34\columnwidth}{!}
{\includegraphics{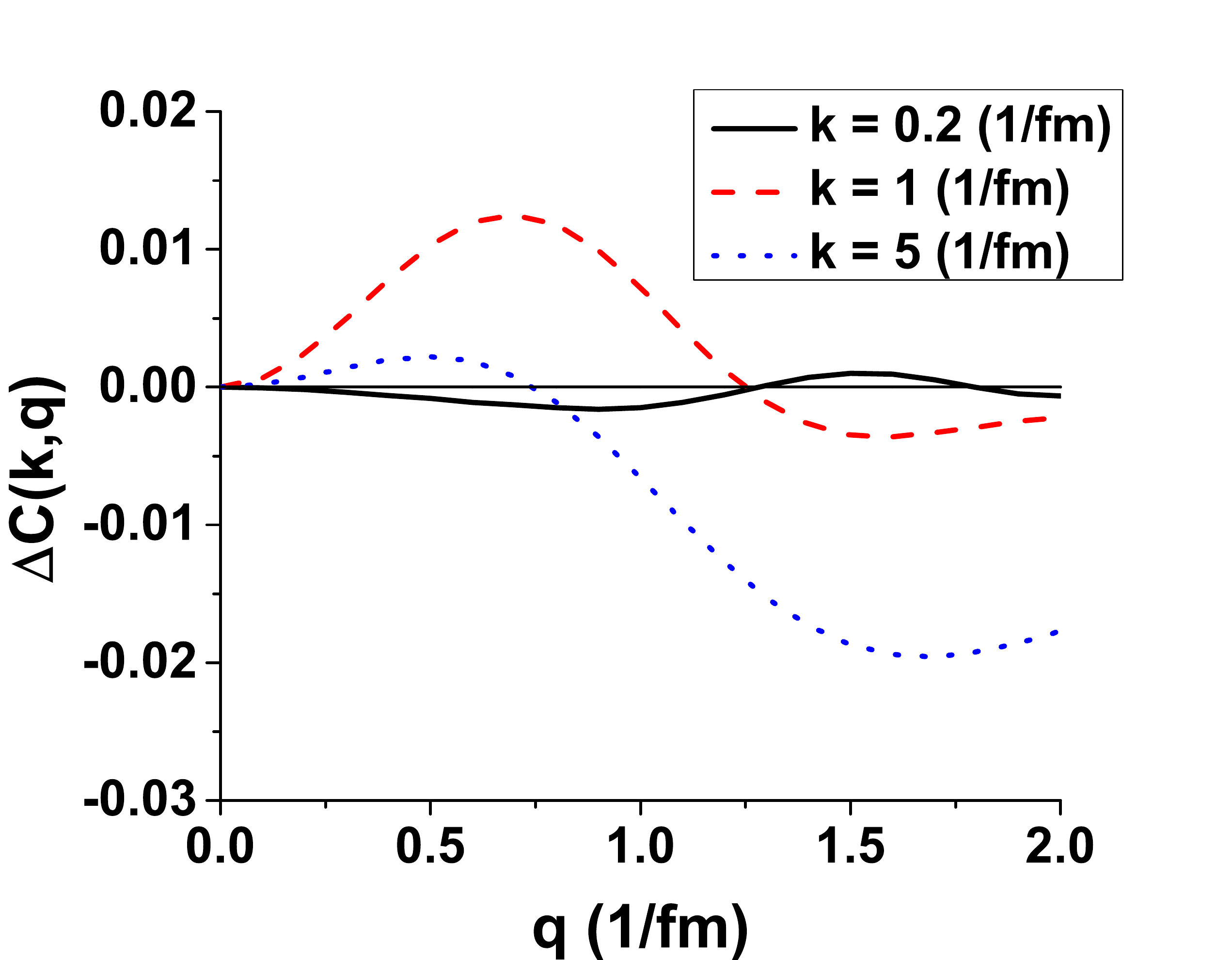}}\!\!\!\!\!\!
\resizebox{0.34\columnwidth}{!}
{\includegraphics{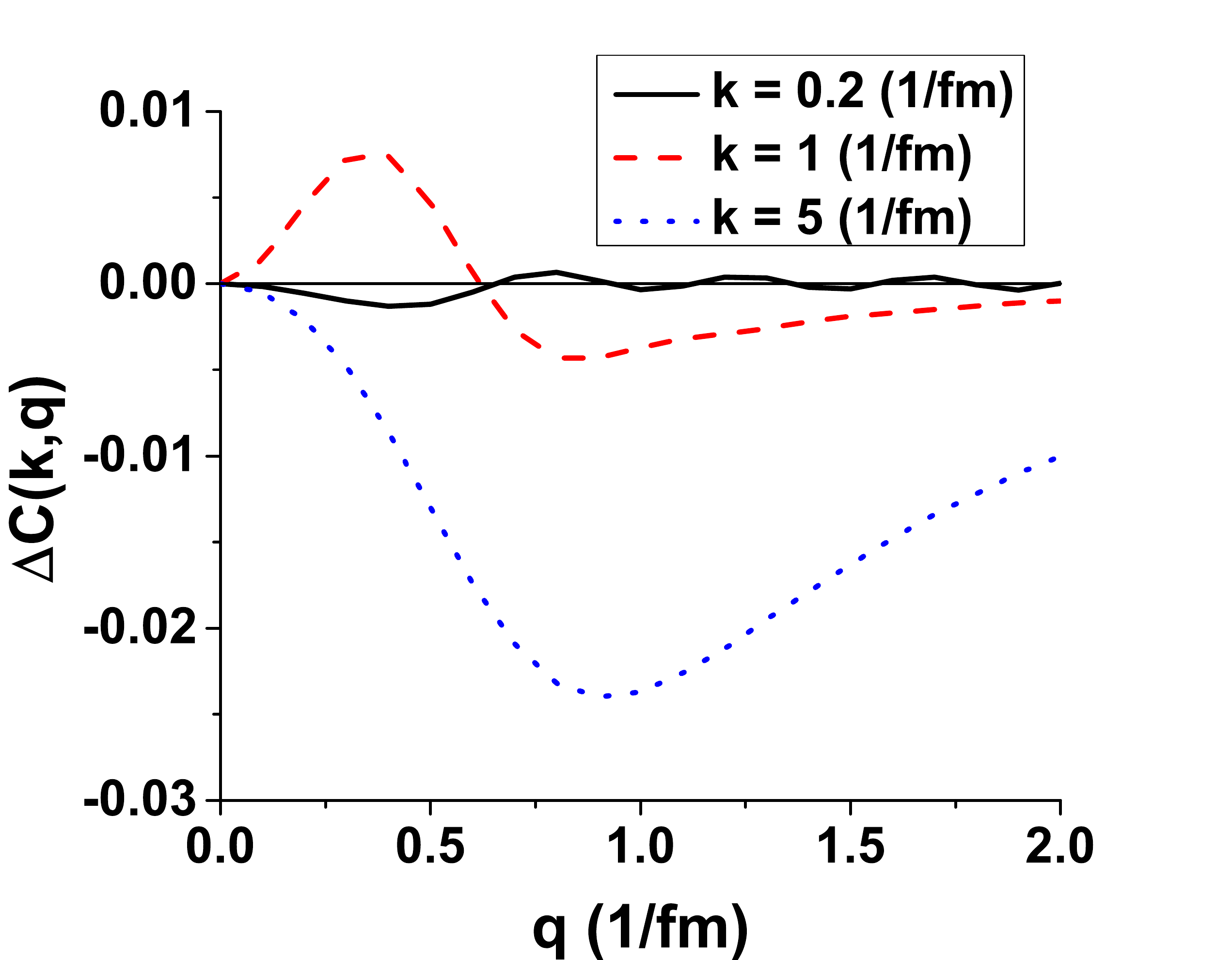}}
\end{center}
\caption{ (color online)
Differential Correlation Function for the exact hydro model 
as function of $q=q_{out}$, with \\
(a) R = 2.500 fm, $\dot{R}$ = 0.250 c, Y = 4.000 fm, $\dot{Y}$ = 0.300 fm, 
$\omega$ = 0.150 c/fm 
at t = 0.0 fm/c. (left figure) \\
(b) R = 3.970 fm, $\dot{R}$ = 0.646 c, Y = 5.258 fm, $\dot{Y}$ = 0.503 fm, 
$\omega$ = 0.059 c/fm 
at t = 3.0 fm/c. (middle figure) \\ 
(c) R = 7.629 fm, $\dot{R}$ = 0.779 c, Y = 8.049 fm, $\dot{Y}$ = 0.591 fm, 
$\omega$ = 0.016 c/fm 
at t = 8.0 fm/c. (right figure) \\
Where the solid black line is for $k=0.2 \, fm^{-1}$, 
the dashed red line is for $k=1 \, fm^{-1}$ 
and the dotted blue line is for $k=5 \, fm^{-1}$.
In (a) the solid black line is close to the axis.}
\label{dcft}
\end{figure}
\end{widetext}

For the initial time, t=0, the DCF is small and positive, 
Figure \ref{dcft} $a$, but
for the later times the amplitude is larger and negative.
Comparing Figures \ref{dcft} $b$ and $c$ we see that as expected 
the peaks of the DCF are more to the left for 
larger size because $R \sim 1/q$ at half width of the correlation function. 
We also see that the amplitude for DCF for Figures \ref{dcft} $b$ and $c$ 
are about the same. As the system continues to increase in size 
we would see a decrease in the amplitude because of the lower $\omega$ value.
For higher values of the temperature we get a smaller amplitude in the DCF, 
or for smaller values of the temperature we get a higher amplitude.
At late times, $t = 8$ fm/c and larger size, the CF is much more 
narrow as shown in Figure \ref{ckqt} $c$. Notice that at low wave numbers,
$k$ the CF has several zero points. This is also reflected in the 
DCF at the same $k$ value, see Figure \ref{dcft} $c$.

\bigskip
\bigskip

The DCF is dependent on the positions of the detectors as 
demonstrated in Figure \ref{dcft}. This dependence can be used to maximize
the amplitude of DCF in a given configuration, based on previous
theoretical estimates. Measuring a single CF at different azimuthal 
angles in the 
plane of rotation does provide the same $C(k,q)$-s as our model is 
azimuthally symmetric! 
Thus, the difference is caused by the measuring two CFs in one event 
with the same source function homogeneity area, but two different
k-vectors at different azimuths with respect to the source area. 
This is also indicated by eq. (27) of \cite{DCF}, where the 
$\epsilon$ sinh$\left(\frac{2 \vec{k} \cdot \vec{u_s}}{T_s}\right)$ 
factor appears, which leads to the sensitivity on vector \vec k.

\section{Conclusion}

The model calculations show that the Differential HBT method can 
give a measure of rotation in this exact hydro model. The differential 
correlation function is dependent on shape, 
temperature, radial velocity and angular velocity. 
Also the detector position is important.

If we eliminate rotation or the radial expansion the DCF vanishes in the model. 
It also indicates that using the estimated small
rotation velocities, $\omega = 0.01 - 0.15 $ c/fm, we get 
a DCF value approaching 2-3 \%.

\section*{Acknowledgements}

This work is supported by the Research Council of Norway, Grant no. 231469.


\end{document}